 \documentclass[preprint]{aastex}
\slugcomment{This version, 2000 Nov. 3}
\newcommand{\kms}{km~s$^{-1}$}

\newcommand{\lam}{$\lambda$}

\slugcomment{This version, 2000 November 10}
\shorttitle{Ultra-high-resolution spectrograph with adaptive optics}
\shortauthors{Ge et al.}
\begin{document}

\title{An Optical Ultra-High-Resolution Cross-dispersed Echelle \\
Spectrograph with Adaptive  Optics\footnote{Observations here were made
at the Starfire Optical Range 1.5 m telescope, operated by the
Directed Energy Directorate, Air Force Research Laboratory}}

\author{Jian Ge\altaffilmark{2}, J. R. P. Angel, B. Jacobsen and N. Woolf}
\affil{Center for Astronomical Adaptive Optics,  University of Arizona,
Tucson, AZ 85721}

\author{R. Q. Fugate}
\affil{Starfire Optical Range,
Air Force Research Laboratory, Kirtland AFB, NM 87117}

\author{J. H. Black}
\affil{{Onsala Space Observatory, Chalmers University of Technology,
S-439 92, Onsala, Sweden}\email{jblack@oso.chalmers.se}}

\and

\author{M. Lloyd-Hart}\affil{Center for Astronomical Adaptive Optics,
University of Arizona, Tucson, AZ 85721}

\altaffiltext{2}{{Permanent Address: Department of Astronomy \&
Astrophysics, Penn State University, University Park, PA 16802}
\email{jian@astro.psu.edu}}

\begin{abstract}
A prototype cross-dispersed optical echelle spectrograph of very high
resolution has been designed and built at Steward Observatory and
tested at the Starfire Optical Range (SOR) 1.5 m telescope.
It is the first spectrograph to take advantage of diffraction-limited
images provided by adaptive optics in order to achieve a potential
resolving power of $R \sim 600,000$.
The wavelength coverage in a single exposure is about 300 \AA, which is
approximately 100 times that of conventional spectrographs operating
at comparable resolution. This was achieved by
recording 60 cross-dispersed orders across the $18\times 18$ mm$^2$
area of the CCD detector. The total efficiency of the
system, including the sky and telescope transmission, spectrograph and
CCD detector, is measured to be 1.3\% at peak, much higher than that
of other ultra-high-resolution spectrographs. Sample stellar spectra
with $R \sim 250,000$ are presented.

\end{abstract}


\section{Introduction}

Ever since a coud\'{e} spectrograph was built for  the Mt.~Wilson 100-inch
telescope (Adams \& Dunham, 1932), it has become clear that the main obstacle
to achieving high spectral resolution  is the loss of light at the entrance
slit. In the traditional seeing-limited domain, the best resolution of
a spectrograph   is coupled  to the  telescope diameter (Schroeder 1987).
Hence, the larger the aperture, the lower the spectral resolution for
a given size of grating.
This coupling limited the best spectral resolution of traditional spectrographs
to $R \sim 30,000$ for 4-m class telescopes (Suntzeff 1995). In order
to obtain higher resolution, all kind of tricks such as image slicers, pupil
slicers, and grating mosaics have been employed, which have resulted in very
large and expensive  spectrographs at the Nasmyth  or Coud\'{e} foci of
telescopes (e.g. Diego  et al. 1995; Vogt et al. 1994; Tull et al. 1994).

Recent  applications of adaptive optics (AO) in astronomy have provided
promising new tools for high resolution spectroscopy. The AO system
corrects the wave-front  distortion caused by atmospheric  turbulence
and delivers diffraction-limited images at the telescope focal plane.
In this AO diffraction-limited domain, the spectrograph size is decoupled
from the size of the telescope aperture, because the image size decreases
in proportion to telescope aperture size.  Therefore, AO enables
ultra-high-resolution spectroscopy with gratings of conventional size.

Ultra-high-resolution spectroscopy with
$R > 500,000$ under seeing-limited conditions has been used in
studies of interstellar and circumstellar matter. It has revealed
small-scale structures and stratification in the interstellar medium and
dynamical evolution of proto-planetary disks (e.g. Lambert et al. 1990;
Crawford et al. 1994a,b; 1997; Crawford 1995; Lauroesch, Meyer, \&\ Blades 2000).
However, the instruments used in these studies
have very low efficiency and small wavelength coverage.
For instance, the total efficiencies, including sky transmission, telescope
transmission, slit loss, spectrograph efficiency and CCD quantum efficiency,
of both the double-pass coud\'e spectrograph ($R = (5 \;{\rm to}\; 6)
\times 10^5$) at the McDonald  Observatory 2.7 m telescope and the
Ultra-High-Resolution Facility (UHRF) ($R = 9.9\times 10^5$) at
the Anglo-Australian Telescope (AAT) are less than 0.5\%, estimated from the
published data.  The wavelength coverage for these ultra-high resolution
spectrographs is only a few \AA\ (Lambert et al. 1990; Hobbs \& Welty 1991;
Diego et al. 1995). The limitations are mainly caused by the special optics
used to achieve great resolving power in these instruments, e.g.
a very narrow slit and double-path optics in the McDonald coud\'e spectrograph
and a confocal image slicer in the UHRF at the AAT.
Therefore, existing ultra-high-resolution spectroscopy
with seeing-limited telescopes is generally restricted to the brightest stars,
and to small sections of the spectrum even with large detectors.
As we will show in this paper, the diffraction-limited images produced by an
AO-corrected telescope can  overcome these limitations.

The advantage of using an adaptive-optics telescope for diffraction-limited
spectroscopy is that high resolution can be achieved without a large
loss of light at the entrance slit of the spectrograph. At visible
wavelengths, the diffraction limit of the telescope (diameter $D=1$ to 4 m)
corresponds to an image size that is typically $r_0/D \sim 1/10$ the
seeing-limited image, where $r_0$ is the  coherence length of the atmosphere,
with a typical value of $r_0\sim 10$ cm
in the V band (Beckers 1993). The very small sizes of diffraction-limited
images make the design of an ultra-high-resolution spectrograph
simple, without image slicers, pupil slicers,
double-pass  mirrors, or grating mosaics.
High-resolution spectrographs can thereby gain substantially in throughput
over conventional spectrographs.  In addition, the smaller image widths
provided by the AO system allow the cross-dispersed orders to be
spaced closer together on two-dimensional photon detectors such as CCDs,
allowing more orders to be observed simultaneously. The total wavelength
coverage can be an order of magnitude larger than that of a conventional
spectrograph.

Spectroscopy at very high resolution, $R = 10^5$ to $10^6$, has important
applications in many areas.  Provided that accurate calibration of
wavelengths and stable line spread functions can be maintained over long time periods, high-resolution
spectroscopy is a powerful technique for detecting
the small, periodic variations in Doppler shift of nearby stars that are
induced by planetary companions. High resolution (and high S/N)
is extremely critical especially for measuring atomic abundances of
rare elements in very metal deficienct, old halo stars, isotopic abundances in
stellar atmospheres and
also abundances of atoms and molecules in
 interstellar matter, which help to trace Galactic evolution.
High resolution spectroscopy provides a radically  improved understanding of the physics of stellar
atmospheres, including effects of convection, line-broadening, pulsations,
and magnetic fields. High resolution spectroscopy  also provides an important tool for investigating
the small-scale structure and turbulence of the interstellar  medium,
and the structures of circumstellar  disks and winds.
In all of these applications, sensitivity (i.e. minimization of light
loss) and large wavelength coverage are crucial: these are exactly the
points on which AO-spectroscopy offers important advantages. The
application of AO to an efficient,
ultra-high-resolution spectrograph with large wavelength coverage will
extend the frontier of high-resolution astronomical spectroscopy to fainter,
more highly reddened sources than have been studied in the past.
This paper presents an  overview of the optical design of the
first ground-based, diffraction-limited, cross-dispersed echelle spectrograph.
It discusses the optical performance of the prototype instrument, shows
some first-light observations at the Starfire
Optical Range (SOR) 1.5 m telescope, and illustrates some special aspects
of using AO for spectroscopy.

\section{OPTICAL DESIGN DETAILS}

\subsection{General Considerations}

Previous observations of interstellar and circumstellar matter show that
significant velocity structures exist at levels corresponding to Doppler
shifts or line-widths of the order of 1 km s$^{-1}$ and that many absorption
line components remain unresolved down to 0.5 km s$^{-1}$
(e.g. Black \& van Dishoeck 1988; Welty et al. 1994). A velocity
resolution of $\sim 1$ km s$^{-1}$ or better
is required to obtain  high-fidelity stellar spectra for studies of
photospheric convection and isotopic abundances, which remain to be
fully explored (e.g. Dravins 1987; Sneden 1995, private communication).
Hence, a resolving power of $R\sim$ 500,000, corresponding to a velocity
resolution $\Delta v = c/R \approx 0.5$ km s$^{-1}$, is an essential
requirement for the ultra-high resolution spectroscopy.

The adaptive optics system at the SOR 1.5 m provides diffraction-limited
images at wavelengths from 0.45 to 1.0 $\mu$m (Fugate et al. 1991).
Therefore, our goals were to  design a cross-dispersed echelle
spectrograph  that takes full advantage of the sharpened
 images in this wavelength range, by achieving a resolving power
of $R = 500,000$,  while squeezing as many echelle orders as possible onto
available CCDs.

When a spectrograph is operated with a diffraction-limited telescope, its
spectral resolving power  depends only on  grating  blaze angle
$\theta_B$, collimator beam size $d$ and wavelength, and is given by
\begin{equation}
R = \frac{2~d~tan~\theta_B}{\lambda},
\end{equation}
for a Littrow arrangement. When matched to the image size
 $\phi \approx \lambda/D$, the entrance slit contains about 50\%\ of the
light from  diffraction-limited images. Thus, we chose a commercially available
Milton Roy R2 echelle grating with ruled area of $242\times 116$ mm$^2$,
groove number density of 23.2 grooves mm$^{-1}$, and
a blaze angle of $63{\fdg}5$ to provide  the main  spectral dispersion.
This echelle provides a diffraction-limited resolving power of
$R= 433,000/\lambda(\mu {\rm m})$, which equals 500,000 at I band, 600,000
at R band and 800,000 at V band. At the same time, it  provides a large number
of relatively short echelle orders at visible wavelengths.
For example,  spectra from 1.0 $\mu$m
to 0.47 $\mu$m  are covered by  87 high echelle orders numbered $m$, from
$m=77$ to $m=163$.

Cross dispersion of these echelle orders can be made by a grating or prism.
In order to  have  uniform order-to-order separations, and high efficiency
for the broad AO corrected wavelengths, and also to make the spectrograph
system more compact, a BK7  prism  with 8$\arcdeg$ apex angle was used in a
double-pass configuration as the cross-disperser.

A Loral $2048\times 2048$ back-illuminated anti-reflection (AR)
 coated CCD with 15 $\mu$m square pixels was envisioned for the initial tests
at the SOR 1.5 m telescope (cf. Lesser 1994).  It was proposed that 3 pixels
(physical size of 45 $\mu$m) should sample the diffraction-limited resolution
element at 0.8 $\mu$m so that a camera of effective focal length $F_C=6$ m
was required. This focal length was derived
from the grating dispersive power of
$\lambda d\beta/d\lambda = 2\tan\theta_B $ and the size of the resolution
element, $\Delta x$, on the detector:
\begin{equation}
R = \frac{\lambda}{\Delta \lambda} = \lambda \frac{d\beta}{d\lambda}\bigl(\frac{F_C}{\Delta x}\bigr)\;\;\;.
\end{equation}
 The resulting focal ratio of the spectrograph is  $f/55$.

The predicted spectral  format for the spectrograph is shown in Figure 1.
The central box is the physical size of the Loral CCD ($30\times 30$ mm$^2$),
which can cover about 90 echelle orders, corresponding to a wavelength range
from $\sim$ 0.45 to 1 $\mu$m, with a spectral coverage of $\sim$ 8 \AA\ per
order. Separations between neighboring orders are expected to be larger than
240 $\mu$m (16 pixels) for all the echelle orders. The full width at half maxium
(FWHM) of each order cross section is about 2 pixels.  The 77th echelle order has
the longest free wavelength range of 320 mm at 1 $\mu$m, which needs 11
continuous settings of the $2048\times 2048$ CCD to cover the whole order.
The $m=170$ echelle
order has the shortest free wavelength range of 140 mm at 0.45 $\mu$m.
The  physical extent of  the free wavelength range of the
grating order $m$ on the image plane was derived from
\begin{equation}
\Delta l(m) = \frac{F_C~\lambda_c}{\sigma {\cos\beta_c}{\cos\gamma}}=
\frac{2~F_C{\tan\theta_B}}{m}\;\;,\;\;\;\gamma =0, \alpha = \beta_c =\theta_B,
\end{equation}
where $\lambda_c$ is the central wavelength of order $m$, $\beta_c$ is the
echelle diffraction angle at the order center,  $\gamma$ is an out-of-plane tilt and $\sigma$ is the grating
constant.

\subsection{Spectrograph Optical Layout}

The main part of the spectrograph layout at the
coud\'e room of the SOR 1.5 m telescope is shown in Figure 2.
A  254 mm diameter folding flat  was used to reduce the overall  length
of the spectrograph so that it could be enclosed within
the covered coud\'e optics bench.  A 254 mm diameter off-axis paraboloid with
a focal length of 6 m was used as the collimator and camera mirrors.  The
paraboloid center is offset 157 mm from the axis to leave enough room for the
entrance slit assembly and the grating-prism box. The  R2 echelle grating is
illuminated in the quasi-Littrow mode in
which an out-of-plane tilt of $\gamma \approx 0.25\arcdeg$ was
introduced so that
a $\sim 45\arcdeg$ folding mirror of $102\times 102$ mm$^2$ could be easily
placed in the output beam to put the object spectra onto the CCD.
This out-of-plane tilt  makes the central wavelength of different echelle
orders slightly shift ($\sim$
0.1\AA), and it  also causes  spectrum lines to be curved (Schroeder
1987). The slope of the curved spectrum lines,
$d\beta/d\gamma  \sim 1\arcdeg $,
and the radius of curvature of
the spectrum, $\rho  \sim 1.5$ m. These effects are so tiny that they
can be easily handled in the data reduction. No measurable degradation
in spectral resolution has been found due to these effects.

The echelle grating is enclosed in a box for protection
and cleanliness. The cross-dispersion double-pass
prism also serves as the entrance window. The physical size of the prism is
$160\times 160$ mm$^2$, with 5 mm thickness at the top and 27 mm
thickness at the bottom. The echelle grating
is supported at three points. The lower two  points are supported through
a pivot, while the top supporting point is connected
to a micrometer which can be precisely dialed to scan along the
echelle orders by changing the incidence and diffraction beam angles.
A spring connected to the back of the grating provides partial support.
The weight of the echelle grating, approximately 1 kg. causes a maximal
distortion of about 1 nm in its shape (Roark \& Young 1975), corresponding
to a 0.02 $\mu$m image shift on the CCD: this is tiny compared
to the pixel size of 15 $\mu$m. The micrometer can
move the incident and diffracted
beam off the blaze angle by $\sim \pm$ 1$\arcdeg$ to
allow complete sampling of the free wavelength range.

Figure 3 shows  spot diagrams on the image plane within a field-of-view (FOV)
of $60\times 60$  mm$^2$
for this spectrograph at different wavelengths. The RMS spot size from
the spectrograph optics is much smaller than  the expected resolution element
of 45 $\mu$m  on the  detector. Therefore the spectrograph resolution
is mainly controlled by the diffraction-limited image size at different
wavelengths.

At the SOR 1.5 m telescope, two different  wavelength bands are available with
AO correction: ``Blue'' leg and ``Red'' legs.
The ``Blue'' leg covers from  0.45 to 0.7 $\mu$m  and the ``Red'' leg  covers
from 0.7 - 1.0 $\mu$m. When photons from one band are used for
scientific observations, photons from  the other band are used for  wavefront
sensing and correction.  The transmission of the telescope/AO system
at the corrected AO focus for both beams
is $\sim$ 20\%. The focal ratios for both beams are $f/107$.
Conversion of the $f/107$ input  beam to the spectrograph $f/55$  beam  was
accomplished by using  two 2.5 cm diameter mirrors and two
anti-reflection-coated achromats: both have diameters of 5 cm and focal
lengths of 200 mm and 400 mm, respectively. We originally proposed to place
the pupil on the grating to avoid vignetting.
However,  this is difficult to achieve with the SOR spectrograph
setup. Nevertheless, due to the slow input beam, there is no
significant vignetting whether or not the pupil is put on the grating.

The spectrograph has  11 reflecting and  10 transmissive surfaces,
including the relay optics before the slit. This leads to large
photon loss. Anti-reflection coating  has been  applied to all
transmissive optics in the spectrograph. The reflection loss is
thereby reduced from its original value of more than 8\%\ to less
than 2\% per surface over the broad wavelength range between
$\sim$ 0.5 to 1.0 $\mu$m. All reflecting  surfaces are
silver-coated to improve the total reflectivity to about 98\% per
reflecting surface. The slit assembly was a customer-made one.
Both slit width and slit height can be adjusted. The slit  can be
dialed to any width between  $\sim$ 30 $\mu$m and $\sim$ 200
$\mu$m with a precision of  $\sim$ 1 $\mu$m. The slit height is
usually set to $\sim$ 100 $\mu$m  to have an optimal throughput
and minimal order overlapping of the cross-dispersed orders.

A reimaging optical system was built to view the reflective
slit jaws.  The guide camera is an Electrim  frame-grabbing
CCD camera with  $972\times 1134$ pixels. Since
both the diffraction-limited images and entrance
slit are relatively small, any small relative motion between them will
cause large photon loss, thus  object acquisition and guide system
become important parts of the AO spectrograph.

Calibration lamp units were developed for calibrating the AO
spectrograph. These include a thorium-argon hollow-cathode lamp
and a tungsten-halogen incandescent lamp. The thorium-argon lamp
provides a large number of narrow emission lines for the
wavelength calibration over the entire working wavelength range,
0.5 to 1.0 $\mu$m, (Palmer \& Engelman 1983). The tungsten  lamp
provides a uniform source for flat-field correction. A
fiber-optics system was carefully built  to feed light beams from
the calibration lamps to the spectrograph in the same way as the
input telescope beam for accurate calibrations.

\section{OPTICAL  PERFORMANCE}

The spectrograph was first tested at the SOR 1.5 m telescope in June 1995,
and was later used for initial  stellar spectroscopy in November 1995
(Ge et al. 1996a,b).  During both runs, most data were
recorded on a SOR Kodak  $2048 \times 2048$ thermo-electrically
cooled  CCD with 9 $\mu$m square pixels, because the prepared Loral
$2048\times 2048$ nitrogen-cooled CCD camera did not work.
The readout noise of the Kodak CCD is about 12 $e^-$
and peak quantum efficiency (QE) is less than 50\%.  In addition, it
generates many hot pixels in a few minute's integration.
Together, the total intrinsic noise from the CCD
itself is typically about 60 $e^-$ for a
ten-minute exposure frame, which prevented us from observing faint targets.
Nevertheless, these observations  demonstrate  that an efficient,
ultra-high-resolution spectrograph with large wavelength coverage
can be built to operate with an AO telescope.

\subsection {Wavelength coverage}

Typical spectra obtained from the Blue and Red legs with
the Kodak CCD are shown in Figures 4a,b. The
Blue leg covers about 60 orders from 0.47 to 0.7 $\mu$m and the
Red leg covers about 30 orders from 0.7 to 1.0 $\mu$m. Figures 5a,b show
different orders in the spatial direction from the two legs.
Cross-dispersed echelle orders are clearly separated in the Blue leg spectrum,
while echelle orders are slightly overlapped
in the Red leg spectrum. These  order overlappings are mainly caused  by the
relatively large diffraction-limited image size at longer  wavelengths and
partially by the scattered light from the uncorrected seeing halo.
The measured separations between neighboring
echelle orders  are about 300 $\mu$m
in the R and I bands and about 360 $\mu$m in the V band.

Figure 6 shows an example of the reduced Vega spectra from the AO Red  leg,
obtained in a ten-minute exposure. The two-dimensional
CCD images were  reduced to one-dimensional spectra  using standard
IRAF routines. The typical signal-to-noise ratio for the reduced
spectra is $\sim$ 200 per pixel.  Many telluric absorption lines are well
resolved and shown in the spectra.  The reduced Red leg spectra cover about
200 \AA\ in one exposure with the
relatively small format Kodak CCD ($18\times 18$ mm$^2$).
The Blue leg  spectra  cover about 300 \AA\ in one exposure with the same
CCD. The reduced spectra are shown in a previous publication (Ge 1998).
 The wavelength coverage achieved with this AO spectrograph already
constitutes an improvement by a
factor of 100 compared to the coverage achieved with the traditional
ultra-high resolution spectrographs (Hobbs \& Welty 1991; Diego et al. 1995).

\subsection {Spectral resolution}

Spectral resolution was measured by a HeNe laser and telluric absorption
lines in stellar spectra. Figure 7 shows a reduced spectrum
of a He-Ne laser with longitudinal mode spacing of 1.09 GHz
taken in a 1 second exposure during the June run in 1995. Clearly,
4 separate laser modes are well resolved by this spectrograph, indicating
a resolving power of about $R = 660,000$ at the wavelength of 0.6328 $\mu$m.
This resolving power is very close to the predicted diffraction-limited
value of $R = 680,000$.
However, the stellar observations of Vega (V = 0.0), $\alpha$ Bootis (V = 0.0)
and $\alpha$ Cygni (V = 1.3) in  longer integrations (e.g. 10 minutes)
give  a much lower resolution of $R \sim$ 250,000 as shown in Figure 8, owing
to the use of a $\sim 100$ $\mu$m  entrance
slit instead of the proposed $\sim 45$ $\mu$m slit.
The slit was widened to accept the dispersed stellar images that are
broadened by the atmospheric differential dispersion during the stellar
observations. The wider slit lowers the spectral
resolving power since the photons from the uncorrected
seeing halos in the AO corrected images filled the slit, unlike the perfect
diffraction-limited HeNe beam (Beckers 1993; Ge et al. 1997; 1998).

The atmospheric differential dispersion provides
relatively large separation among images from different wavelengths
covered by our spectrograph.
For  example, compared to the reference wavelength of 1 $\mu$m,
the atmospheric  dispersion at 45$\arcdeg$ elevation at SOR was
$\sim 0{\farcs}1$ in the I band, $\sim 0{\farcs}3$ in the R band, and
$\sim 0{\farcs}6$ in the V band during the run,  when
the atmospheric pressure  was about 610 mm Hg and the temperature was about
$20\degr$ C  and the water vapor pressure was  about 8 mm Hg
(cf. Filippenko 1982). The atmospheric dispersion direction at the SOR 1.5 m
was perpendicular to the entrance slit height direction.
The relative atmospheric  differential dispersion was about  $0{\farcs}2$
to $0{\farcs}3$ for both the Blue and  Red legs.  The SOR 1.5-m telescope had no atmospheric dispersion corrector at the
time of these observations. Therefore, it was
impossible to apply a  slit with a $0{\farcs}1$ width for either of the two AO
bands without dispersion correction.
However, a $0{\farcs}1$ slit is necessary to  provide a resolving power of
$ R > 500,000$.   Therefore,  we compromised by opening the slit to
accommodate the dispersed stellar images to $\sim 0{\farcs}3$, which
corresponds to 100 $\mu$m at the detector and which yields $R \sim 250,000$.

\subsection {Detection efficiency}

The total system detection efficiency was measured
during both runs in 1995.
The total efficiency of the spectrograph in the November run
increased by a factor of
$\sim 3$ compared to that in the June run  after
all transmissive optics surfaces were anti-reflection coated
and all mirror surfaces were coated with protective silver.  For example,
the peak efficiency for most orders increased to about 0.8\% from
0.3\% in the  June  run.
Figure 9 shows the total efficiency measured at different echelle orders
around the peaks of their blaze functions. The overall contour of the
total efficiency versus wavelength is similar to that of the SOR
telescope/AO transmission*QE curve except near the edges of both wavelength
bands. The difference was probably caused by the atmospheric differential
dispersion of stellar images, which most severely affected the extreme
wavelengths of the Blue and Red AO bands. For comparison, the total efficiency
of this spectrograph is about a  factor of 3 higher than that of the AAT UHRF
(Diego et al. 1995). Furthermore, considering the fact that the QE of the
Kodak CCD was only about half that of the proposed Loral CCD, another factor
of 2 gain in the total efficiency could be achieved.

We have also measured the blaze function of each order, which shows a similar
shape for all echelle orders. Figure 10 is an example of a
measured grating blaze function at order $m=121$.
The central peak efficiency, about 62\%, is approximately twice the
efficiency at either end of the free wavelength range determined by
$\lambda/m$.
The unsymmetrical shape of the blaze function was caused by
the slightly different cross-section of the echelle grating  dialed
to different wavelength along the dispersion order.

\subsection{Spectrograph stability and scattered light performance}

We have also tested the stability of the spectrograph, which is important
in studies of the small line shifts in the stellar spectra that are caused by
convection,  or by gravitational perturbations due to companions, or by
stellar activity probed via asteroseismology (e.g. Trimble 1995).
This property was tested  by performing a cross-correlation analysis between
reduced one-dimensional spectra from
two sequential observation frames with identical settings and physical
conditions. Despite  a 6 m focal length of the spectrograph,
the shift between neighboring  frames is small.
For instance, the typical relative shift is about 0.17 pixel over a
one-hour period, which corresponds to a characteristic Doppler velocity
shift of 37 m s$^{-1}$.  This demonstrates good short-term stability of the
spectrograph. We found the heat generated by the thermally cooled CCD
and the residual turbulence inside the Coud\'e bench caused the image drift.
Further improvement is possible by keeping the instrument
 temperature constant and the enclosure isolated from the rest of the Coud\'e
 environment.

For an  astronomical measurement,  scattered light could
limit the precision of the results. A direct measure of the scattered light
in the echelle spectrograph was obtained from the $\zeta$ Per spectrum  in
the region of the telluric O$_2$ A-band. The interorder background
has been subtracted by the IRAF BACKGROUND program. We compare the A-band
absorption lines in the $\zeta$ Per spectrum to those of the solar-flux
atlas of Kurucz et al. (1984).
The solar spectrum was smoothed approximately to our resolution via
the boxcar technique of IRAF. Figure 11 shows the fit
of the smoothed solar telluric A-band spectrum to the $\zeta$ Per spectrum.
Uncertainties from the continuum fitting are responsible for the difference
between the two non-saturated parts of the spectra.
The heavily saturated parts from both spectra are indistinguishable,
demonstrating that the scattered light was tiny and negligible in our
spectrograph.

\subsection {Observations of interstellar clouds at $R \sim 250,000$}

During the November run,  some atomic
absorption lines from the diffuse clouds toward nearby
bright stars were observed with the AO spectrograph, demonstrating the
scientific capability of this instrument. The observed targets include
$\zeta$ Persei (V = 2.8) and $\alpha$ Cygni (V = 1.3). The typical exposure times
for $\zeta$ Persei and $\alpha$ Cygni is 10 min and 5 min, repectively.
The typical S/N in the reduced spectra is $\sim$ 200 per pixel for $\zeta$ Persei
and $\sim$ 300 per pixel for $\alpha$ Cygni.

Figure 12 shows the observed
profiles of interstellar absorption lines of K I at $\lambda= 7698.974$ \AA\
toward these stars. Five velocity components from the interstellar clouds
toward $\alpha$ Cyg were identified at heliocentric
velocities of $-$22.46 \kms, $-$13.74 \kms, $-$8.91 \kms, $-$3.79 \kms\
and 0.51 \kms, and most of these are well separated.
These identifications are in harmony with the components observed by
Welty et al. (1994) in the Na I D$_1$ line profiles.
For the interstellar clouds toward $\zeta$ Per, the K I absorption lines  are
blended with each other. We therefore applied a profile-fitting
program from the IRAF to discern the different individual interstellar clouds
that contribute to the absorption line profiles. Six components at heliocentric
velocities  10.29 \kms, 12.00 \kms, 13.17 \kms,
14.57\kms, 16.17 \kms\ and 17.18 \kms, were found, which provide the best
and unique fit to the line profile. These are also
consistent with the  Na I D$_1$ line profile measurements of Welty
et al. (1994).

The high sensitivity of our new instrument allows us to combine
the K I measurements with previous results  to extend the relationship
between the Na I and K I abundances to lower levels of potassium
than have been studied before.
Figure 13 shows the relationship between Na I and K I abundance in the
Milky Way diffuse clouds. At high abundances, the results are from the previous
measurements of Hobbs (1976) with a high-resolution PEPSIOS spectrometer
at $\sim$ 1 \kms\ resolution. At low abundances, the results are derived
from the Na I data of Welty et al. (1994) together with our
measurements of K I in the November run.  The extended relation is consistent
with a constant ratio of column densities $N({\rm Na})/N({\rm K})=60$;
however,  the new measurements at low column densities exhibit much larger
scatter than the previous data at high column densities.
This scatter cannot be simply
explained  by a normal photonization model, which  predicts the
correlation between  $N({\rm Na})$ and $N({\rm K})$ (Hobbs 1976 and
references therein). The scatter could also be caused  by the
inhomogeneous distribution of neutral Na and K atoms in the small diffuse
clouds, where the neutral fractions of these easily ionized elements are
sensitive to the local densities of electrons and of UV starlight. Where
small clouds merge together to form stronger absorptions (i.e. larger diffuse
clouds), the abundance inhomogeneity is averaged out to show less scatter
in the larger diffuse clouds. The new results are based only on the diffuse
clouds toward $\alpha$ Cygni: future observations toward other directions
are needed  to confirm this result.

The observations of the narrow and weak interstellar
K I lines  toward $\alpha$ Cygni have also provided another
measurement of the spectrograph resolving power. The
average FWHM of these absorption lines is  1.31 \kms, while the average
intrinsic Doppler $b$-value of the same clouds observed in Na I is
$b({\rm Na\ I}) = 0.47$ \kms\ (Welty et al. 1994). In the limit that the
intrinsic width is due to thermal Doppler broadening, the corresponding
intrinsic width expected in potassium is  $b({\rm K\ I}) = 0.36$
\kms. The deduced resolution of the instrument is $\Delta V = 1.16$ \kms,
or $R\approx 260,000$, in harmony with the estimate based on the profiles
of telluric absorption lines in the Vega spectrum (Figure 8).

The sensitivity of high-resolution spectra to narrow, weak absorption
features is also limited by the signal/noise ratio, $S/N$, that can be
achieved in reasonable integration times and by the shape of the
instrumental response function. Despite the large number of transmissive
and reflective optical elements in the spectrograph, the profiles of
narrow absorption lines appear gaussian down to levels of approximately
3\% of the peak absorption. This has been estimated from close examination
of the profile shapes of the interstellar K I line and of nearby telluric
O$_2$ lines in the spectrum of $\zeta$ Per. The demonstrated short-term
stability of the spectrograph (\S 3.4) ensures that the instrumental
profile can be measured and the observed spectra de-convolved in cases
where the detailed shapes of far-wing profiles might be important.

The interstellar spectra also provide a good test of the sensitivity in
terms of minimum measurable equivalent width of absorption features.  The
spectrum of $\zeta$ Per covered the wavelengths of several other potentially
interesting interstellar features. In particular, interstellar Rb I
($\lambda = 7800.27$ \AA) has been sought by several previous investigators
because it is an important probe of stellar nucleosynthesis and mixing.
Although Jura \&\ Smith (1981) reported the discovery of interstellar Rb I with
a line of equivalent width $W_{\lambda}=2.5\pm 0.8$ m\AA\ toward
$\zeta$ Oph, subsequent observations of $\zeta$ Oph,
$\zeta$ Per, and o Per revealed no line stronger than $W_{\lambda}
=1.5$ m\AA\ (3$\sigma$) in any of those directions (Federman et al.
1985). Our spectrum of
$\zeta$ Per at $\lambda\approx 7800$ \AA\ shows no significant absorption
feature. We find an upper limit to the equivalent width of a narrow
interstellar Rb I line of $W_{\lambda}\leq 1.1$ m\AA (3$\sigma$). This
corresponds to $S/N\approx 80$ per resolution element at the wavelength
of interest. Our limit is very similar to that published by Federman
et al. (1985), who used a larger telescope (2.7 m) at lower resolution.
It is clear that the present limit could be easily improved with a
more efficient detector and longer integration times.

We also searched for a few lines of the A--X (3,0) band of interstellar
C$_2$ in absorption toward $\zeta$ Per. Two weak lines of telluric
O$_2$ are found at 7720.3 and 7721.5 \AA, with equivalent widths of
$W_{\lambda}=0.7$ and 1.0 m\AA, respectively. The noise level in this
part of the spectrum corresponds to a limiting equivalent width
$W_{\lambda}\leq 0.5$ m\AA (3$\sigma$). No interstellar C$_2$ lines
are apparent above this level. This is not too surprising given the
difficulty of detecting the (3,0) band toward any star (van Dishoeck
\&\ Black 1986) and the weakness of the lines of the stronger (2,0) band
toward $\zeta$ Per (Chaffee et al. 1980). Although no new physical information
can be derived from these unsuccessful searches for Rb I and C$_2$, the
data clearly show that the first test measurements already achieved
a level of sensitivity comparable to that of the best previous observations
of $\zeta$ Per.

It is also worth noting that the weak telluric O$_2$ lines at 7700--7800
\AA\  can provide alternative wavelength calibration. The wavelengths of
carefully measured O$_2$ lines are consistent with the calibration
based on Th-Ar lamp spectra within $\pm 0.5$ km s$^{-1}$ in Doppler velocity.
In fact, the absolute wavelengths of O$_2$ lines in this wavelength
region are still uncertain at this level (Brown \&\ Plymate 2000).

\section{SPECIAL ASPECTS OF USING AO FOR SPECTROSCOPY}

The benefit of combining AO with spectroscopy is that the
diffraction core of the image is small both in the dispersion and spatial
direction. This is what allows the entire spectrum to be recorded on a
reasonable area of CCD. This benefit, which may be as much as a factor of
100  or
more in recorded spectrum, must be weighed against losses, which though
small compared to the benefit are nonetheless significant, and affect
how an AO spectrograph is used.

The image provided by an AO telescope is not just a diffraction-limited
image. There is a diffraction-limited core sitting on a broad
pedestal whose size is the seeing width (Beckers 1993; Ge et al. 1997, 1998).
The image is characterized by its
Strehl Ratio ($SR$), the fractional decrease in the central intensity from a
perfect diffraction-limited image. Mar\'{e}chal approximation gives
\begin{equation}
{SR} \approx \exp(-\Delta)~,
\end{equation}
 where
\begin{equation}
\Delta \propto \frac{D^{5/6} \sec^2~Z}{\lambda^2},
\end{equation}
$D$ is the aperture diameter of a telescope, and $Z$ is the zenith distance
(Beckers, 1993).
A $SR$ of 50\% is typical for the 1.5 m telescope at 0.8 $\mu$m
when observing near zenith. The Strehl Ratio gets poorer as the telescope
moves away from the zenith, so that for example at $30\arcdeg$ elevation
(an airmass of 2), this Strehl Ratio would drop to about 6\% at the same
wavelength. For this reason the
AO system is very rarely used below an elevation angle of
45$\arcdeg$ where  the corresponding Strehl Ratio is 25\%.

Because of partial diffraction limit of the AO corrected images, the slit
width becomes more critical for the final resolution and throughput.
A wide slit width corresponding to 2.44 $\lambda/D$, the diameter of the first
Airy dark ring, will admit about 84\% of the energy from the
diffraction-limited core into the spectrograph, but it can only provide
resolution of about
$0.8 d\tan\theta_B/\lambda$ because the broad background of the seeing halo
smooths out the higher resolution provided only by diffraction. While a
narrow slit width of about $\lambda/D$ does provide about a factor of two
higher resolution, $2 d\tan\theta_B/\lambda$, approximately half of the energy
from the diffraction-limited core is lost due to slit diffraction and slit
loss itself. Thus, the final correct balance will depend on the scientific
goal: higher throughput but lower resolution versus higher resolution but
lower throughput (see Ge et al. 1997 for a detailed  study).

AO has so far been largely used for imaging where the Strehl Ratio has
not been very critical, and where it has been possible to work with a
relatively feeble core of an image, just as in the early days of the
Hubble Space Telescope. Now with  AO spectroscopy, the Strehl Ratio
becomes very important. And the Strehl Ratio available
depends both on the seeing and on the
AO design. This will set a practical limit to the wavelength
range of the spectrograph. Table 1 shows how the Strehl Ratio at different
wavelengths is likely to change with the seeing. It is seen that the efficiency
is likely to  be quite low at wavelengths  $\lambda\leq 0.5$ $\mu$m, and so a
practical lower limit to the usable wavelength at the AO-corrected telescope
is likely to be about 0.5 $\mu$m. The maximum efficiency (including
$\sim$ 20\% telescope and AO transmission, $\sim$ 50\% slit transmission,
$\sim$ 50\% spectrograph and  $\sim$ 90\%
QE CCD) of  $\sim$ 4\% for the existing SOR 1.5 m AO telescope and the AO
spetrograph
can be potentially reached  in the I band with a $\sim$ 2$\lambda/D$ slit.
Future improvement in the telescope and AO transmission can increase the
total efficiency further.

Another aspect in which AO spectroscopy differs from traditional
spectroscopy is in the effect of atmospheric dispersion. Atmospheric dispersion
is independent of image size. For a conventional $1\arcsec$ image, a dispersion
of about $1\arcsec$ is noticeable, which is not very important. Instead
with AO, the dispersion becomes very important. Thus, for example,
at 1 $\mu$m, the diffraction FWHM at the 1.5 m telescope is $0{\farcs}13$. At
the altitude of SOR and at $45\arcdeg$ elevation, the dispersion from the
image position  is given in Table 2.
It can be seen that the observations at SOR 1.5 m telescope were skating on
the edge of being impossible without correction for atmospheric dispersion.
An optical device, such as a Risley prism pair, can be built to correct
atmospheric dispersion for AO spectroscopy.

In conclusion, the AO spectrograph has some features that need to
be factored into its performance and operation that are less than obvious.
The range of elevation of sources must be reduced.
In particular, the observing program needs to be tightly tailored to when
stars are nearest to zenith, and objects at lower elevations should be
observed when the seeing is best and the atmospheric motions are slow.
Correction for atmospheric dispersion will become a crucial part of observing.
Guiding needs to be done on the image at the slit, not at the telescope.
Inter-order spacing needs to allow for diffraction at the longer wavelengths.
Resolution and throughput of the spectrograph have to be carefully
balanced by the entrance slit width.


\acknowledgements
We thank J. Christou, D. Barnaby and other staff of the SOR adaptive optics
group for their great help during our observation runs. We are  grateful to
M. Lesser, C. Corbally, D. Baxter, and L. Ulrickson for preparing and lending
us the $2048\times 2048$ Loral CCDs;  G. Schmidt, J. Hill, and B. Martin for
lending us some  optics; and D. Ouellette, B. McMillan, R. Kurucz, C. Sneden,
D. Meyer, and S. Federman for useful discussions. This work was supported by
NSF grant AST-9421311. NSO/Kitt Peak FTS data used here were produced by
NSO/NOAO. Research in atomic and molecular astrophysics at Onsala Space
Observatory is supported by the Swedish Natural Sciences Research Council
and the Swedish National Space Board.

\clearpage

\begin{figure}
\plotone{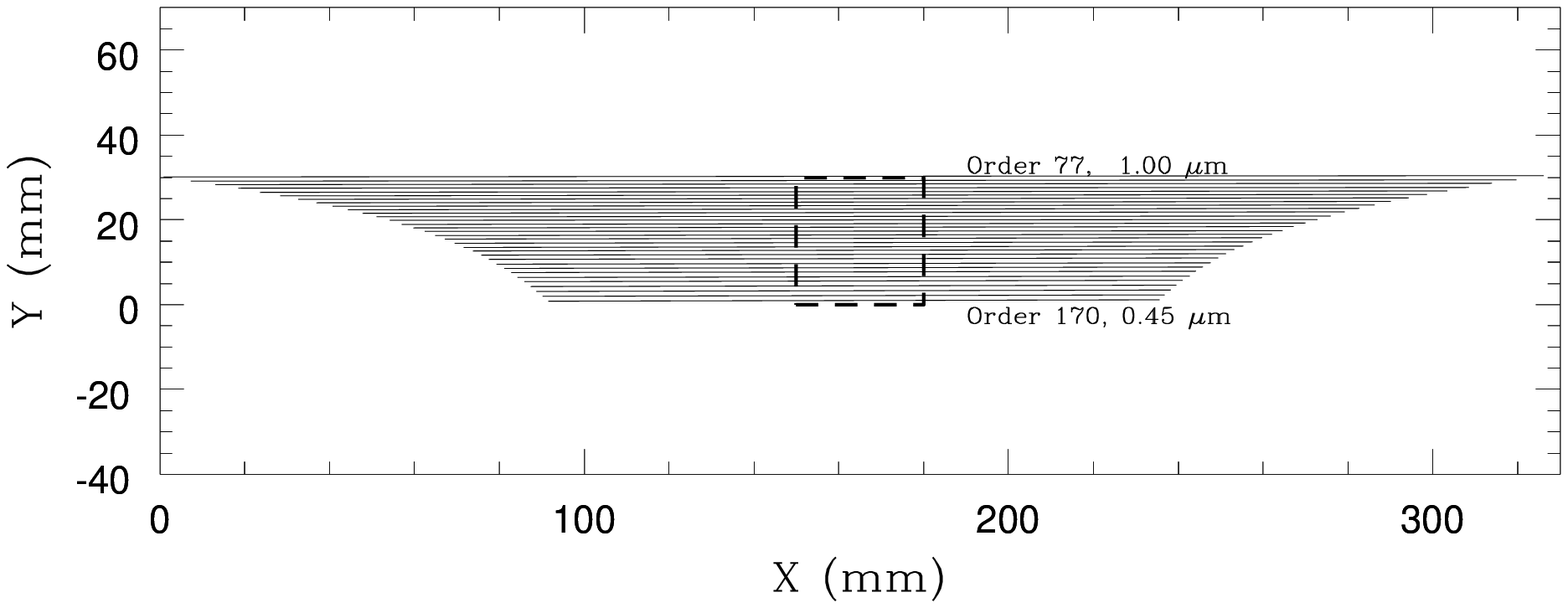}
\caption{ CCD echellogram obtained with the
 R-2 echelle grating in the Littrow configuration
and the  BK7 thin prism used in double-pass mode. The focal length of the
collimator/camera was 6 m. The X- and Y-axes are  in units of millimeters.
Wavelength increases from left to right and from bottom to top.
The dashed box represents the physical size of a $2048\times
2048$ CCD with 15 $\mu$m pixels.\label{fig1}}
\end{figure}


\clearpage

\begin{figure}
\plotone{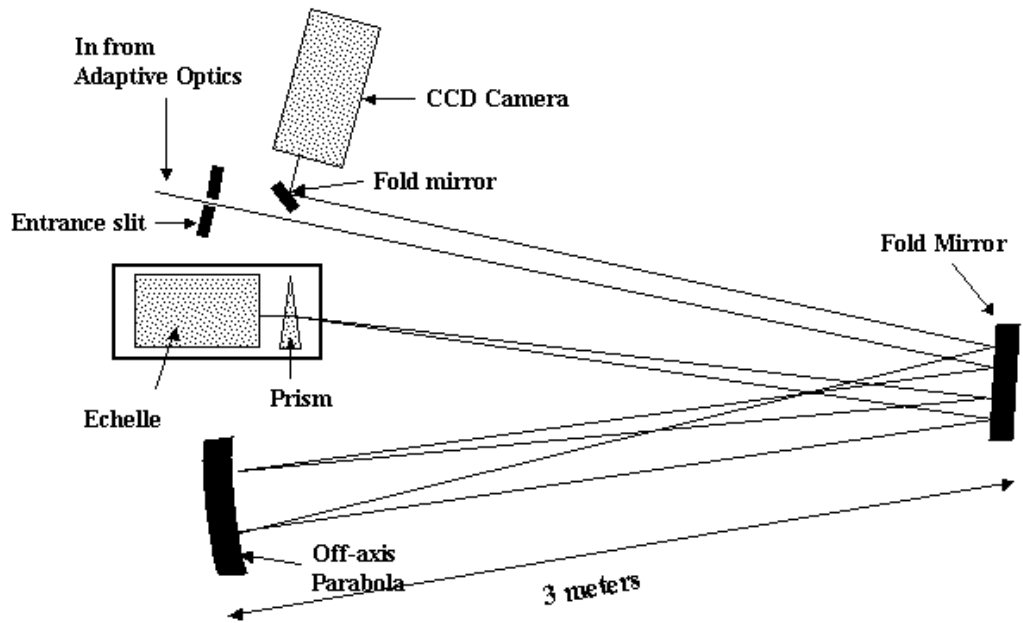}
\caption{Optical layout of the cross-dispersed echelle spectrograph at the
SOR 1.5 m telescope. \label{fig2}}
\end{figure}

\clearpage

\begin{figure}
\plotone{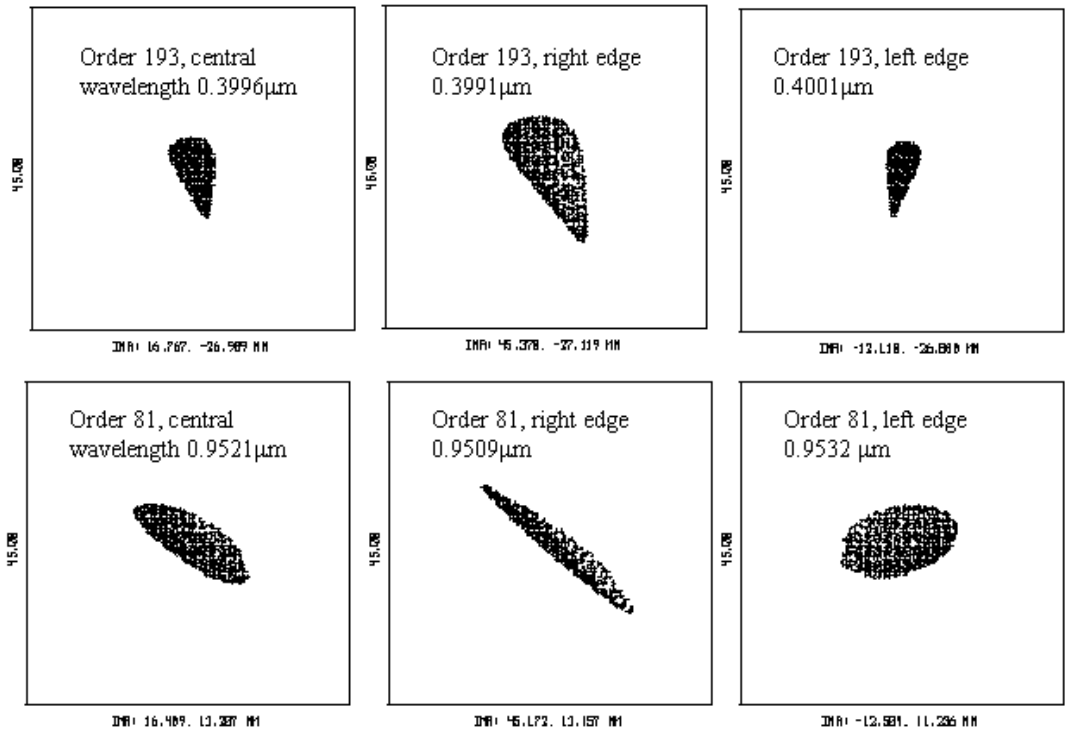} \figcaption{Spot diagrams within a $60\times
60$ mm$^2$ field of view on the image plane for the optical design
of the spectrograph, including relay systems before the entrance
slit. The square box size is 45 $\mu$m. The RMS image diameters
are about 6 $\mu$m in the wavelength range from 0.5 $\mu$m to 1
$\mu$m. \label{fig3}}
\end{figure}

\clearpage

\begin{figure}
\plotone{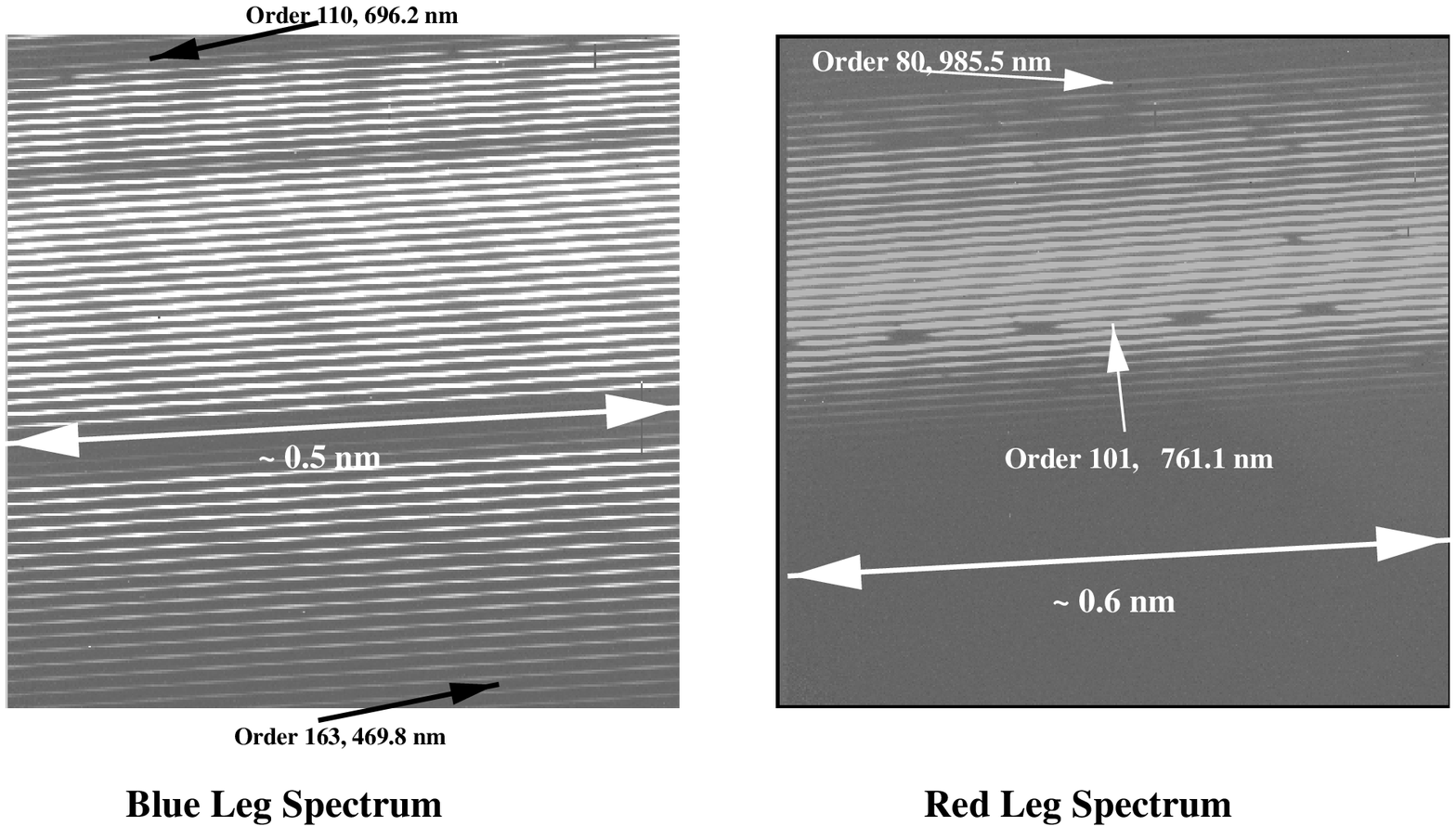}
\caption{(a). Vega spectrum obtained from the blue leg of the SOR 1.5 m
telescope/AO system with the Kodak 2k$\times$2k CCD. Approximately 60
echelle orders are covered, corresponding to a wavelength of approximately
0.47 $\mu$m at the bottom  and 0.7$\mu$m
at the top. (b). Vega spectrum from the red leg with the same Kodak CCD.
The 30 orders cover from  0.7 $\mu$m at the bottom to
1.0 $\mu$m at the top.\label{fig4}}
\end{figure}

\clearpage

\begin{figure}
\plotone{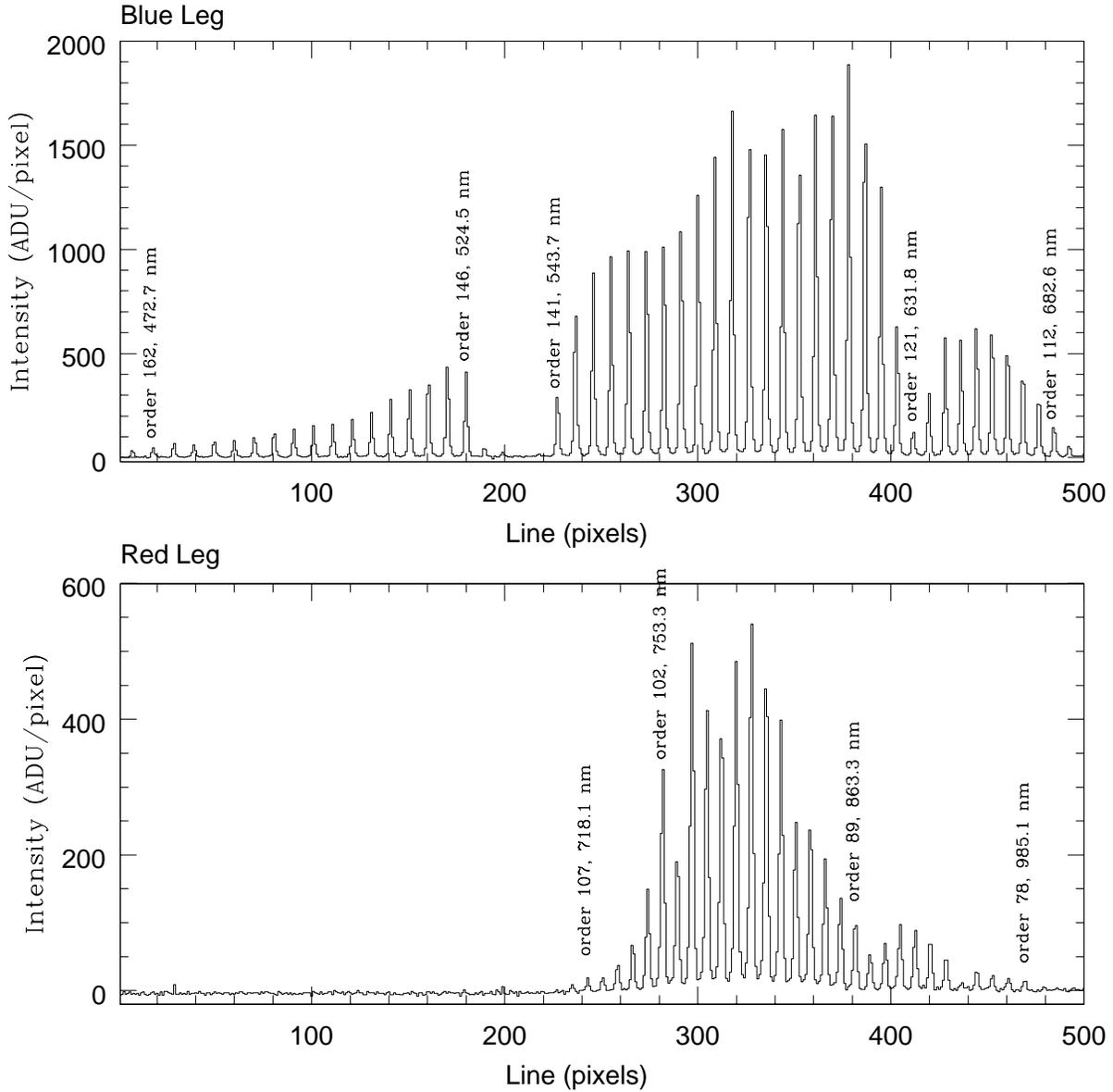}
\caption{(a). Tracing across the echelle orders in the blue
spectrum of Vega. The left side corresponds to the short  wavelength. The
missing orders in the middle regions are caused by a dichroic in the
adaptive optics (Fugate 1995, Private communication).
The rapid decrease of intensity in the short wavelength range
is probably caused by the atmospheric differential dispersion, which
displaces some photons at short wavelengths outside the slit.
(b). Tracing across the orders in the red spectrum of Vega. The left is
the short-wavelength side. The reduced sensitivity of the CCD at the longer
wavelengths is probably responsible for the
long wavelength limit. The short wavelength  limit is set by beamsplitters in
the AO system. \label{fig5}}
\end{figure}

\clearpage

\begin{figure}
\plotone{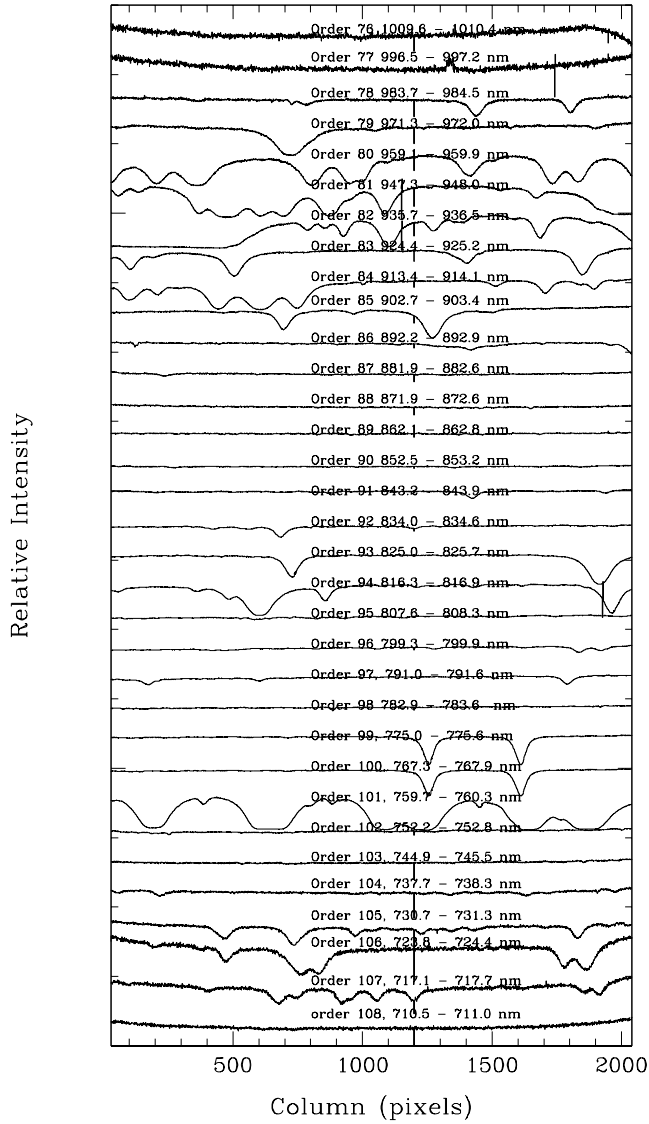}
\caption{The reduced Vega spectra from the red
leg, which covers from 710 nm to 1010 nm. The intensity of each spectrum
 is rescaled to
a unit value and  also offset by
$\sim$ 0.5 from each other.  The 2k$\times$2k Kodak CCD
covers 200 \AA\ in one exposure with the AO spectrograph. \label{fig6}}
\end{figure}

\clearpage

\begin{figure}
\plotone{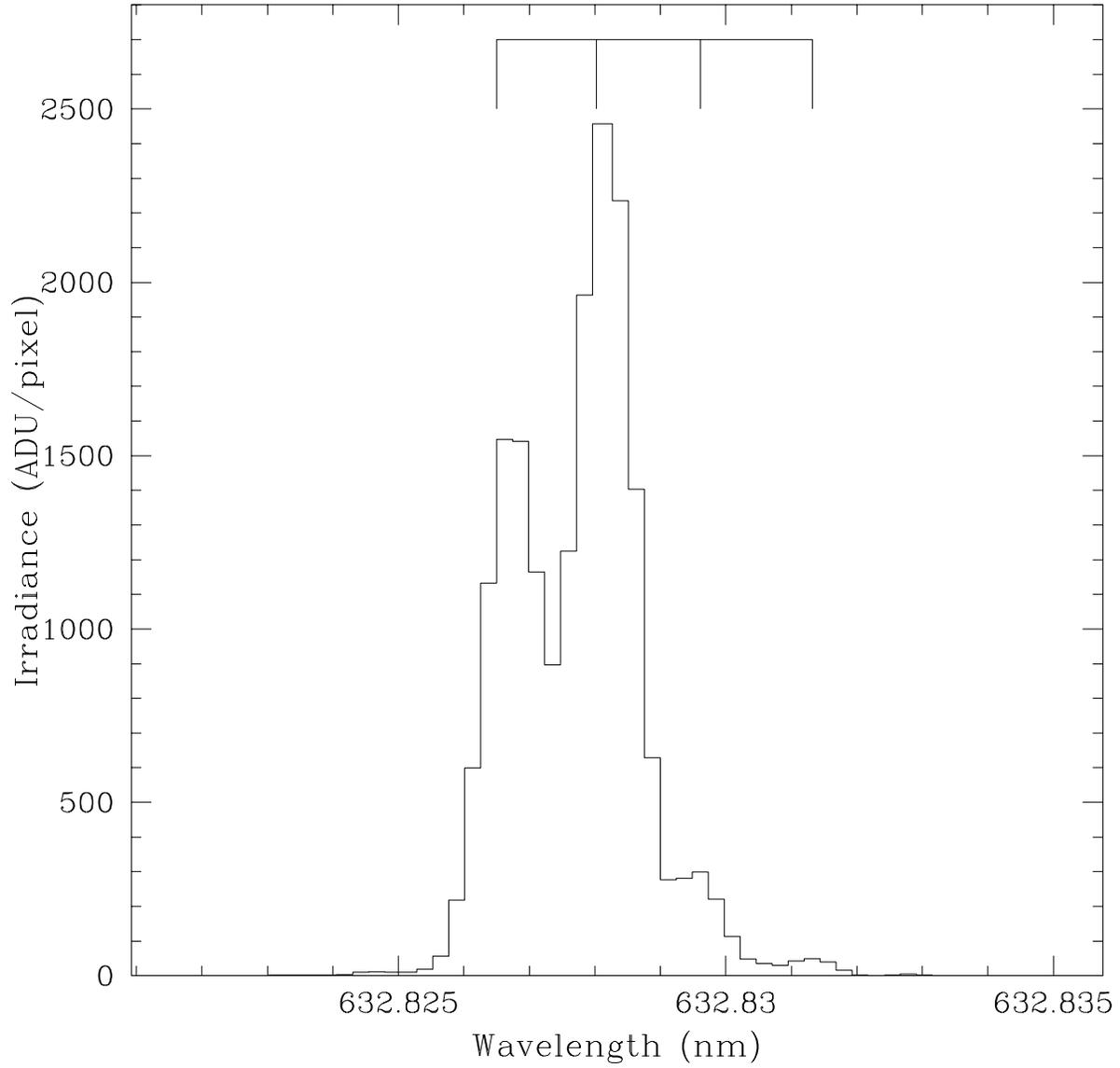}
\caption{The spectral profile of the He-Ne laser at 6328.160 \AA, where
4 adjacent laser modes separated by 0.0618 \AA\ are clearly resolved. The
FWHM resolution is 3.9 pixels or 36 $\mu$m, corresponding to a resolving
power of 660,000. \label{fig7}}
\end{figure}

\clearpage


\clearpage

\begin{figure}
\plotone{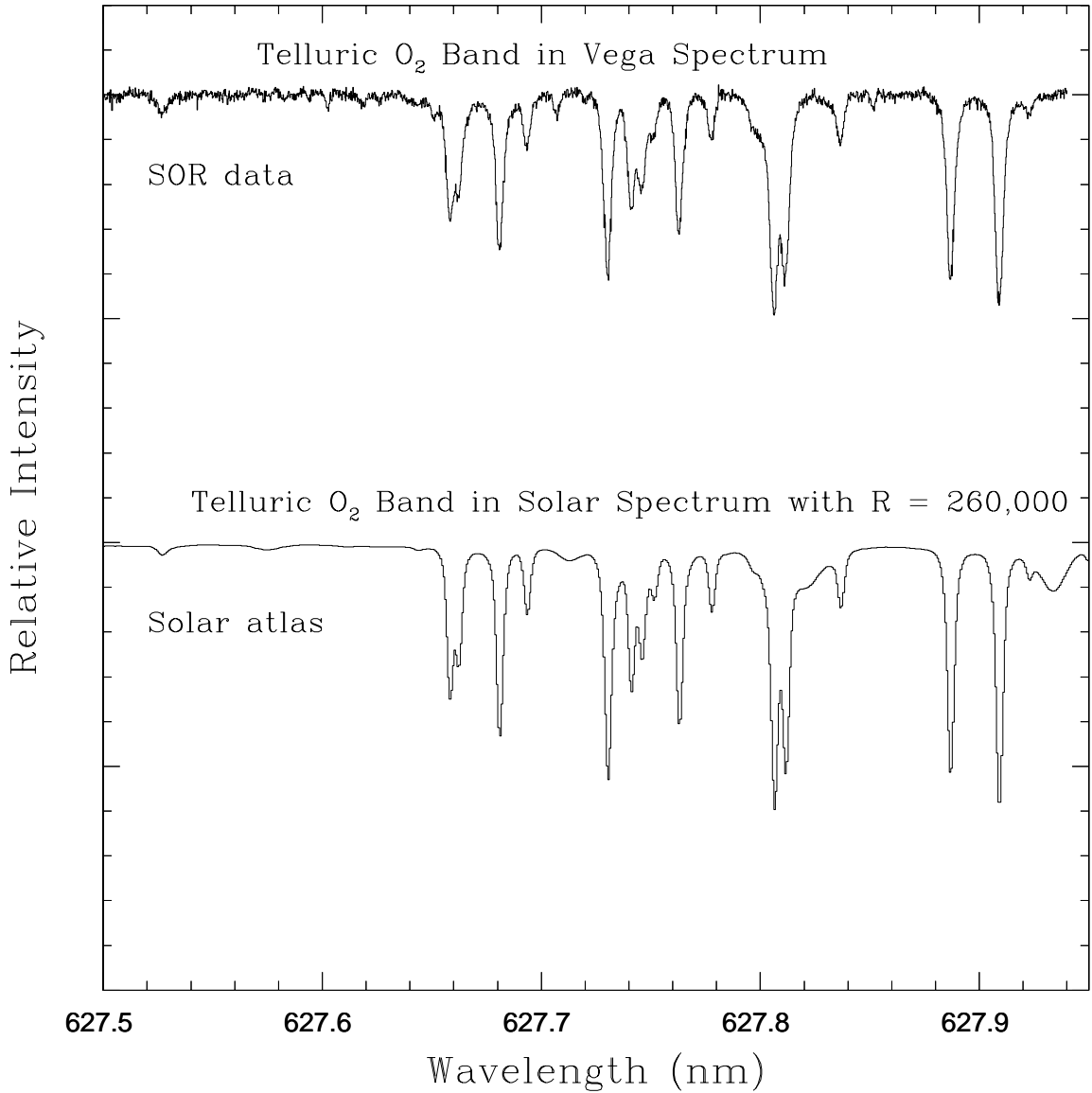}
\caption{Part of the telluric O$_2$ absorption lines   in the
 Vega spectrum obtained with the SOR/AO echelle  spectrograph.
The Kurucz  et al. (1984) solar spectrum is shown for comparison, after
smoothing to a resolution of $R =$ 260,000 via convolution with a Gaussian
function.  Two sets of telluric lines show similar line profiles, indicating
a spectral resolving power of $R \approx$ 250,000 for the AO spectrograph.
\label{fig8}}
\end{figure}

\clearpage

\begin{figure}
\plotone{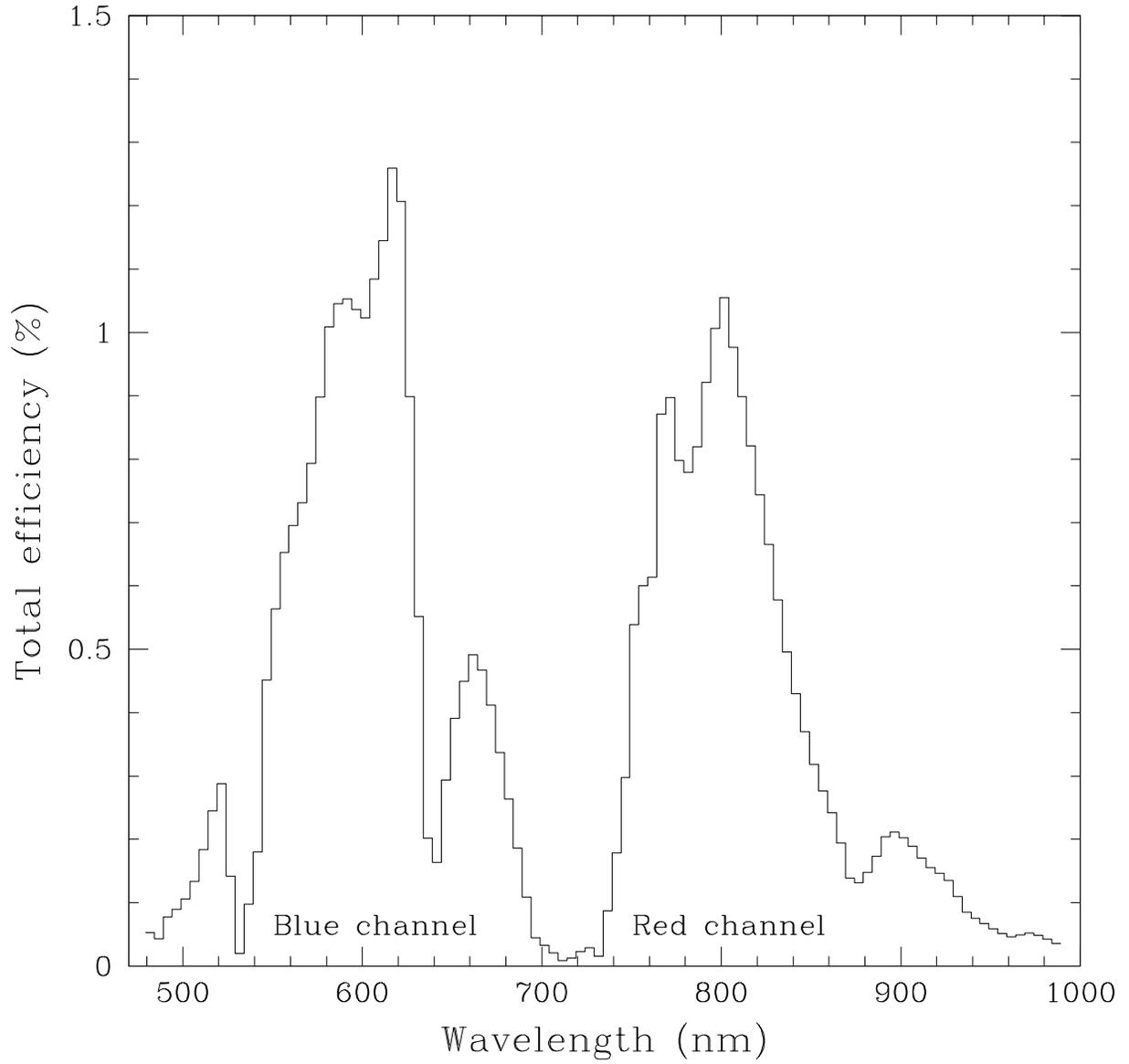}
\caption{The measured total detection efficiency of the SOR/AO echelle
spectrograph as a  function of wavelength, which includes the sky
transmission, telescope/AO transmission, Strehl ratio of the image,
slit loss, and the combined efficiency of the
spectrograph and Kodak CCD. \label{fig9}}
\end{figure}

\clearpage

\begin{figure}
\plotone{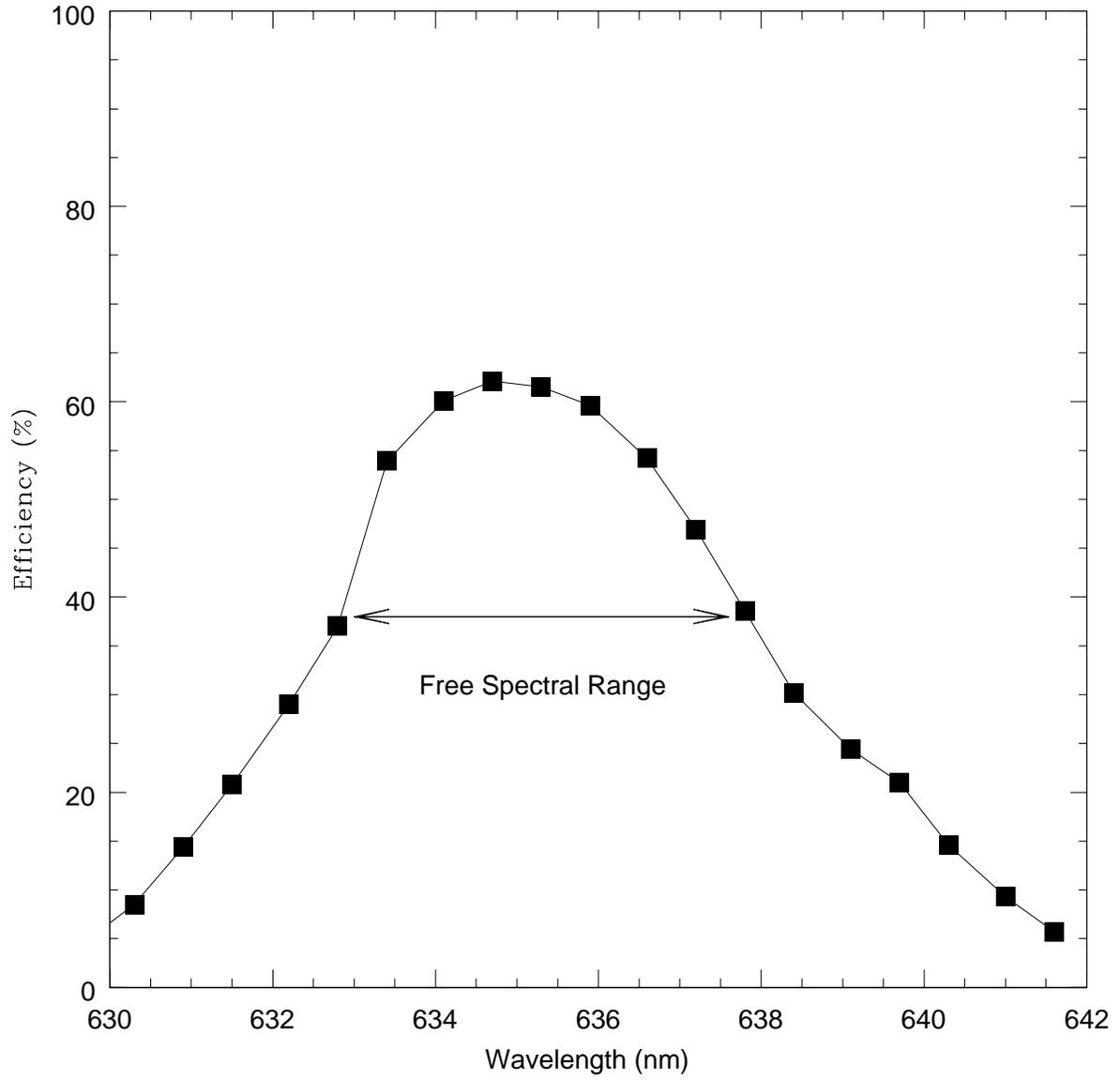}
\caption{An example of a measured blaze function of
the echelle at order 121, where the central wavelength is
$\lambda_c$ = 6353 \AA. \label{fig10}}
\end{figure}

\clearpage

\begin{figure}
\plotone{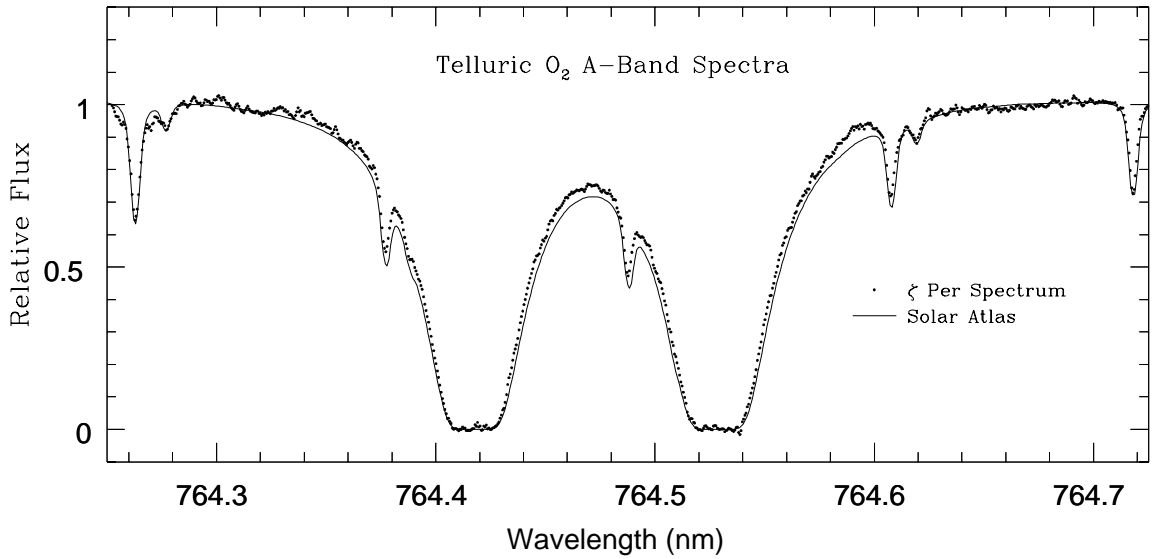}
\caption{Part of the telluric O$_2$ A-band absorption lines from the
$\zeta$ Per spectrum, compared to the Kurucz et al. (1984) solar atlas
spectrum. Uncertainties from the spectral normalization are probably
responsible  for the difference between these two spectra. The deepest
parts of the line profiles are virtually indistinguishable in these
spectra, indicating that
scattered light was not a problem for the SOR/AO echelle spectrograph.
\label{fig11}}
\end{figure}

\clearpage

\begin{figure}
\plotone{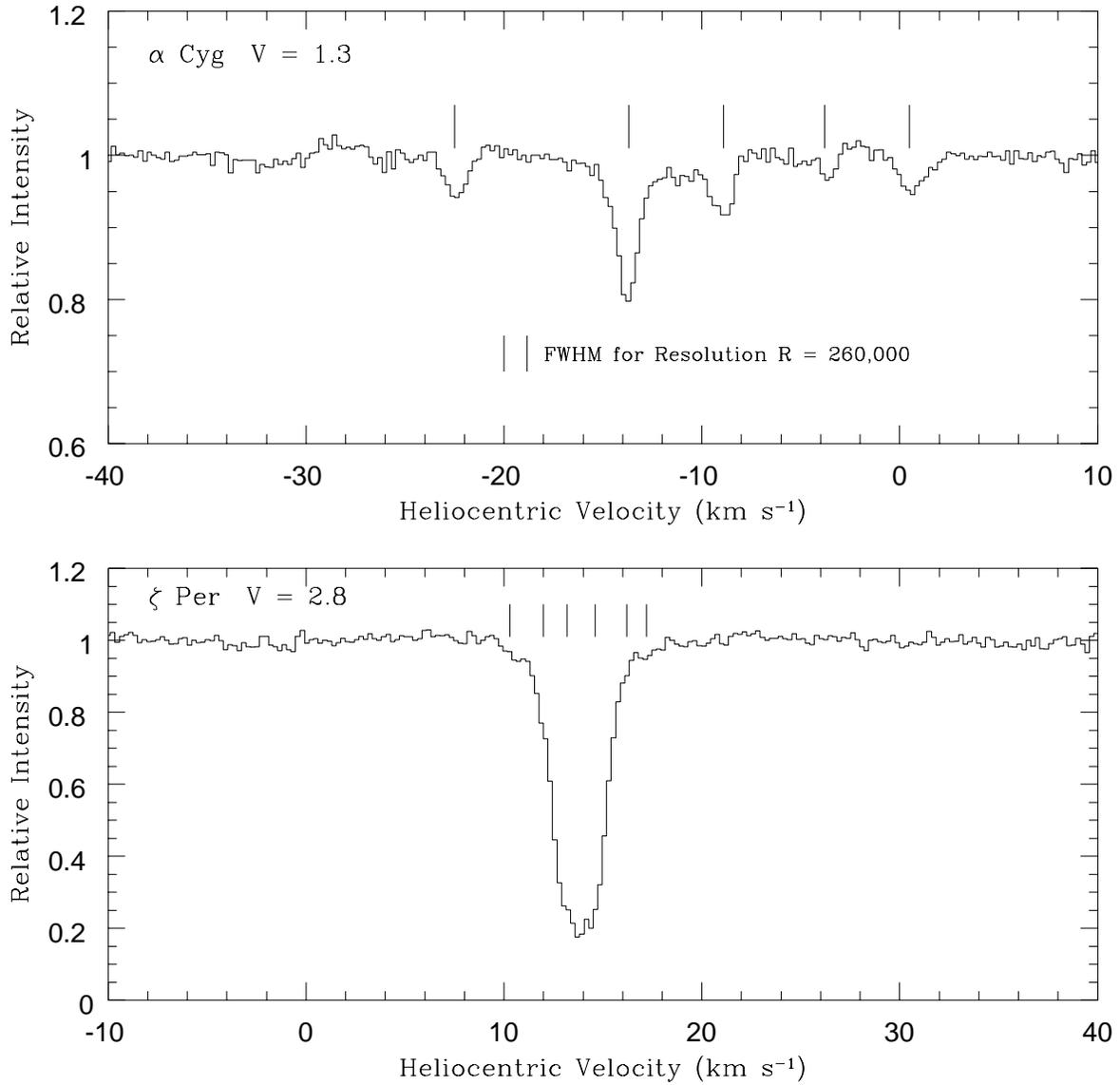}
\caption{The interstellar K I \lam\ 7698 \AA\ absorption lines
 in the spectra of (a) $\alpha$ Cyg and (b) $\zeta$ Per. Tic marks indicate
 location of each velocity components in the best fitting. \label{fig12}}
 \end{figure}

\begin{figure}
\plotone{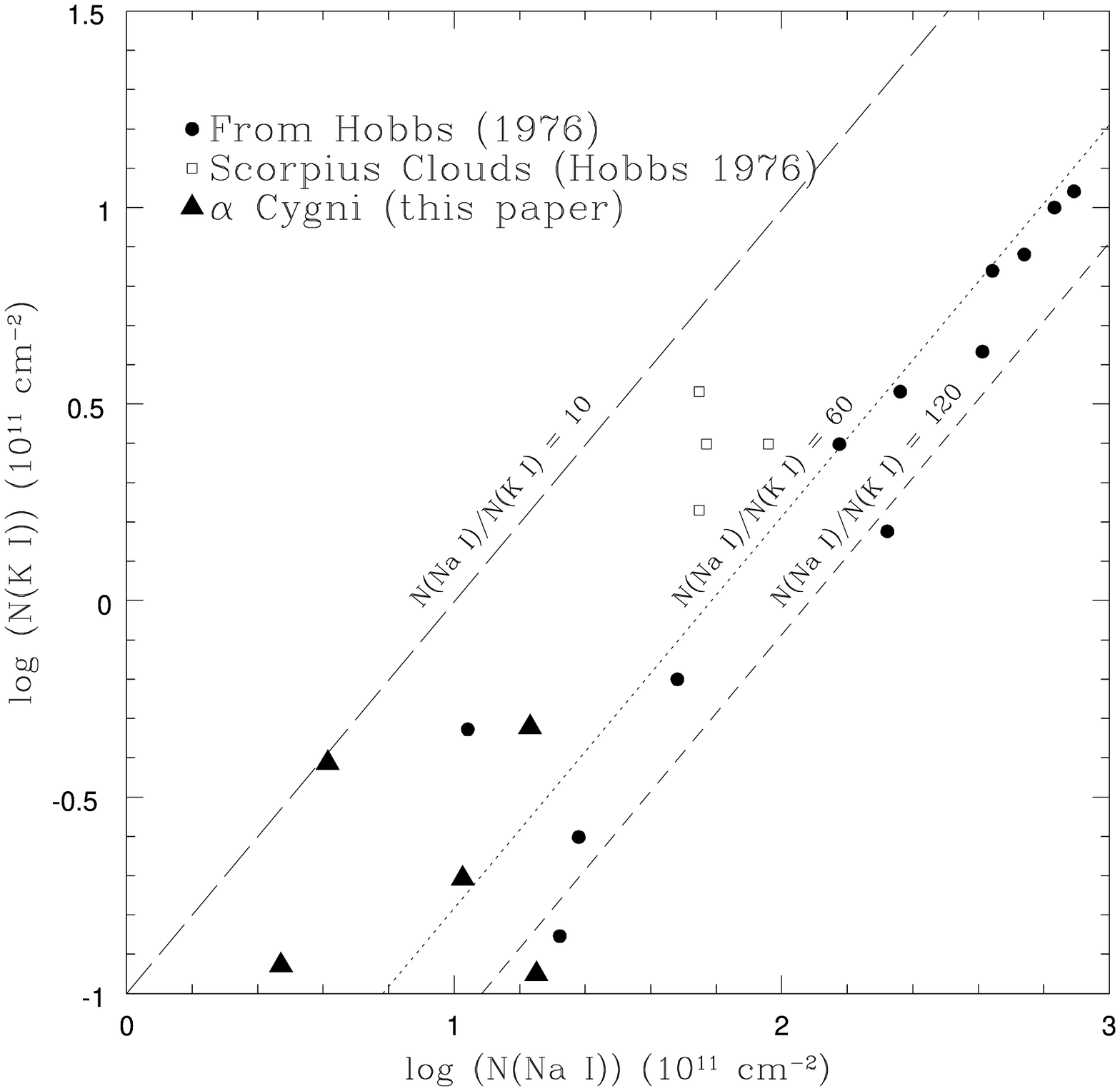}
\caption{The correlation of Na I and K I  column density in the Milky
May diffuse  clouds. \label{fig13}}
\end{figure}

\clearpage
\footnotesize
\begin{flushleft}
Table 1. Strehl Ratio of the AO Corrected Images vs Wavelength at
the SOR 1.5 m Telescope under different atmospheric conditions at
SOR$^a$

\end{flushleft}
\begin{tabular}{ccccc} \hline
\hline
    1.0 $\mu$m  &    0.84 $\mu$m  & 0.707 $\mu$m &  0.599 $\mu$  & 0.500$\mu$\\
\hline
70\%&61\%&49\%&37\%&24\%\\
61\%&49\%&37\%&24\%&14\%\\
49\%&37\%&24\%&14\%&6\%\\
37\%&24\%&14\%&6\%&2\%\\
\hline
\end{tabular}

\noindent $^a$ Typical atmospheric turbulence parameters at 0.55
$\mu$m  are $r_0 \sim 6$ cm and the  Greenwood frequency  $\sim$
45 Hz, which corresponds to a coherence time $t_0$ = 3 ms (see Ge
1998 for details).

\begin{flushleft}
Table 2. Atmospheric Differential Dispersion at 45$^\circ$ elevation at SOR$^a$
\end{flushleft}
\begin{tabular}{cccccccccc} \hline
\hline
1.0$\mu$m&0.90$\mu$m&0.80$\mu$m&0.75$\mu$m&0.70$\mu$m&0.65$\mu$m&0.60$\mu$m&0.55$\mu$&0.50$\mu$m&0.45$\mu$m\\
0.0&0.06$''$&0.15$''$&0.21$''$&0.27$''$&0.36$''$&0.48$''$&0.62$''$&0.82$''$&1.09$''$\\
\hline
\end{tabular}
\smallskip

\noindent $^a$ The calculation was based on P = 610 mm Hg, T = 10 $^\circ$C
and water vapor pressure 8 mm Hg (Filippenko 1982).


\end{document}